\documentclass[pdftex,twocolumn,epjc3]{svjour3}          

\RequirePackage[T1]{fontenc}

\RequirePackage{graphicx}
\RequirePackage{mathptmx}      
\RequirePackage{flushend}
\RequirePackage[numbers,sort&compress]{natbib}
\RequirePackage[colorlinks,citecolor=blue,urlcolor=blue,linkcolor=blue]{hyperref}
\usepackage{xcolor}
\usepackage{hyperref}
\usepackage{amsfonts}
\usepackage{amsmath}
\usepackage{multirow}
\usepackage{setspace}
\usepackage{amssymb}

\def \ee{\end{equation}}
\def \be{\begin{equation}}
\def \bea{\begin{eqnarray}}
\def \eea{\end{eqnarray}}

\begin{document}

\title{Scale-dependent planar Anti-de Sitter black hole}


\author{\'Angel Rinc\'on\thanksref{eA,addrPUC}
        \and
        Ernesto Contreras\thanksref{eE,addrESPOL}
        \and
        Pedro Bargue\~no\thanksref{eP,addrUA}
        \and
        Benjamin Koch\thanksref{eB,addrPUC}
}

\thankstext{eA}{e-mail: arrincon@uc.cl}
\thankstext{eP}{e-mail: p.bargueno@uniandes.edu.co}
\thankstext{eE}{e-mail: econtreras@yachaytech.edu.ec}
\thankstext{eB}{e-mail: bkoch@fis.puc.cl}

\institute{Instituto de F{\'i}sica, Pontificia Universidad Cat{\'o}lica de Chile,\\ Av. Vicu{\~n}a Mackenna 4860, Santiago, Chile.\label{addrPUC}
          \and
          Departamento de F\'{\i}sica, Universidad de los Andes, Cra.1E
No.18A-10, Bogot\'a, Colombia.\label{addrUA}
          \and
Yachay Tech University, School of Physical Sciences \& Nanotechnology, 100119-Urcuqu\'i, Ecuador .\label{addrESPOL}
}


\maketitle

\begin{abstract}
In this work, we investigate four-dimensional planar black hole solutions in anti-de Sitter spacetimes in light of 
the so-called scale-dependent scenario. To obtain this new family of solutions, the classical couplings of the theory, i.e., 
the gravitational coupling $G_0$ and the cosmological constant $\Lambda_0$, are not taken to be fixed values anymore. 
Thus, those classical parameters evolve to functions which change along the ``height'' coordinate, $z$. 
The effective Einstein field equations are solved, and the results are analyzed and compared with the classical counterpart. 
Finally, some thermodynamic properties of the presented scale--dependent black hole are investigated.
\end{abstract}

\section{Introduction}


%
Although a consistent formulation of quantum gravity remains an open task, there are
several promising approaches in this direction.
Even though those candidate theories differ in their approach, their variables, and techniques,
they have a usefull common feature.
Their low energy effective action for the gravitational field acquires a scale dependence. 
This is observed through the coupling 
constants which evolve from constant values to scale--dependent functions with respect to certain energy scale.
Similar approaches have been considered before, but the motivation and implementation in those approaches is quite different. This is the case of 
the Brans--Dicke (BD) theory \cite{Brans:1961sx}, which treats the Newton coupling constant as an auxiliary scalar field. 
Thus, adopting this formalism, the link between $G$ and $\phi$ is just $\phi \rightarrow G^{-1}$ which means that the 
Einstein coupling constant takes the equivalent form $\kappa \equiv 8 \pi \phi^{-1}$. This deviation from the classical 
Einstein Gravity take into account that Newton coupling could be a field and not a fixed value. Despite of it, BD theory is 
still a classical theory and it does not include the possibility for the other parameters included in the action to evolve to 
scale--dependent functions.
What is more, it is very--well known that an effective description takes the effective action as a functional whose 
coefficients show a scale dependence, which is a generic result of quantum field theory.

In this sense, the aforementioned effective action $\Gamma[g_{\mu \nu}, k]$ contains a set of couplings inherited from the classical theory but incorporating the scale
dependence, where $k$ stands for an undetermined 
scale--dependence.
 Specifically, $\{G_k, (\cdots)_k\}$ comes from $\{G_0, (\cdots)_0\}$ (where $(\cdots)$ denotes any other
coupling present in the theory).
The probably most successful implementation of those ideas was achieved within the so called
Asymptotic Safety (AS) program, where a non-trivial ultra violet fixed point for the leading dimensionless
gravitational couplings was conjectured \cite{Weinberg:1976xy} and  found 
\cite{Wetterich:1992yh,Morris:1993qb,Reuter:1996cp,Reuter:2001ag,Litim:2002xm,Litim:2003vp,Niedermaier:2006wt,Niedermaier:2006ns,Gies:2006wv,Machado:2007ea,Percacci:2007sz,Codello:2008vh,Benedetti:2009rx,Manrique:2009uh,Manrique:2010am,Manrique:2010mq,Eichhorn:2010tb,Litim:2011cp,Falls:2013bv,Dona:2013qba,Falls:2014tra,Eichhorn:2018yfc,Eichhorn:2017egq}.

Recently, scale--dependent gravity has
been used to construct black hole backgrounds both by improving classical solutions with the scale dependent couplings from AS \cite{Bonanno:1998ye,Bonanno:2000ep,Emoto:2005te,Bonanno:2006eu,Reuter:2006rg,Koch:2007yt,Hewett:2007st,Litim:2007iu,Burschil:2009va,Falls:2010he,Casadio:2010fw,Reuter:2010xb,Cai:2010zh,Falls:2012nd,Becker:2012js,Koch:2013owa,Koch:2014cqa,Gonzalez:2015upa,Torres:2017ygl,Pawlowski:2018swz}
and by solving the  gap equations of a generic scale dependent action 
\cite{Contreras:2013hua,Koch:2013rwa,Rodrigues:2015hba,Koch:2015nva,Koch:2016uso,Rincon:2017goj,Rincon:2017ypd,Rincon:2018sgd,Contreras:2018dhs,Rincon:2018dsq,Contreras:2018gct,Contreras:2018gpl,Rincon:2018lyd,Contreras:2018swc}.
Even more, regular black holes \cite{Contreras:2017eza} and traversable (vacuum) wormholes \cite{Contreras:2018swc} have been shown to exist within this approach. In this sense, scale--dependent gravity might shed light on how to cure, in an effective way, some of the classical problems
which appear in classical general relativity. From the cosmological side, 
the impact of scale dependence has been explored in various ways
\cite{Bonanno:2001xi,Weinberg:2009wa,Tye:2010an,Bonanno:2010bt,Bonanno:2001hi,Koch:2010nn,Grande:2011xf,Copeland:2013vva,Bonanno:2015fga,Bonanno:2017gji,Hernandez-Arboleda:2018qdo,Bonanno:2018gck,Canales:2018tbn}.

It is important to note that almost all the exact black hole solutions found in the context of scale--dependent
gravity (but the cosmological and a rotating scale--dependent BTZ black hole 
which has been recently reported \cite{Rincon:2018lyd} belong to the spherically symmetric case. 
Therefore, the role of different geometries for scale--dependent black hole solutions (if any), remains to
be investigated. 
This is the purpose of the present work, with emphasis in planar black hole geometries. 
Although this
work could be easily extended to the the toroidal case, we do not expect substantial differences with respect to the
spherically symmetric case. 
On the contrary, the planar nature of the scale-dependent black hole we will present in the
present work makes it an ideal candidate to see the effects of scale dependence when a non--compact event horizon is present.

The manuscript is organized as follows. In Sect. \ref{classAdS} we review the main aspects of the
classical planar AdS black hole solution. Section \ref{scale_setting} is devoted to introduce the 
scale--dependent model. In sections \ref{BlackHoleSolution} and  \ref{IT} we obtain the 
scale--dependent solution and study their geometrical and thermodynamical aspects. Some final comments 
are given in the last section.

\section{Classical planar Anti-de Sitter theory and black hole solution}\label{classAdS}
The Einstein-Hilbert action is, in four dimensions, given by
\begin{align}\label{classical_action}
I_0[g_{\mu\nu}] &= \int \mathrm{d}^{4}x \sqrt{-g}
\bigg[
\frac{1}{2 \kappa_0} 
\bigg(R - 2\Lambda_0 \bigg)
\bigg],
\end{align}
where $\kappa_0 \equiv 8 \pi G_0$ is the gravitational coupling, $G_0$ is
Newton's constant, $\Lambda_0$ is the cosmological coupling, $g$ is the determinant of the metric and $R$ the Ricci scalar. 
In what follow we assume that the space-time is plane-symmetric and time-independent. Besides, we assume the coordinate set $x^{\mu} = \{ t, x, y, z \}$, we use the metric signature $(-, +, +, +)$, and natural units $(c = \hbar= k_B = 1)$ such that the action is dimensionless. The line element is then written according to
\begin{align}
\mathrm{d}s^2 &= -f_0(z) \mathrm{d}t^2 + f_0(z)^{-1}\mathrm{d}z^2 + (L z)^2(\mathrm{d}x^2 + \mathrm{d}y^2).
\end{align}
Please, note that the term $L z$ is dimensionless. In addition, it is remarkable that the cosmological coupling is usually related to $L$ by $ 3 L^2 \equiv -\Lambda_0 > 0 $ (where $\Lambda_0$ denotes the negative cosmological constant). This constraint is, however, relaxed in order to obtain a more general set of solutions.
To be consistent with the classical scale setting, we take $G_0=1$. Varying the classical action \ref{classical_action} yields the equations of motion, i.e., 
\begin{align}
	R_{\mu \nu} - \frac{1}{2}R g_{\mu \nu} &= -\Lambda_0 g_{\mu \nu}.
\end{align}
For a vacuum solution we only have the cosmological constant contribution, and the lapse function becomes 
%
\begin{align}\label{f0cl}
f_0(z) &= - \frac{1}{3}\Lambda_0 z^2 -\frac{4 M_0}{L z} 
\end{align}
or, in terms of the event horizon, $z_0$, we have
\begin{align}
f_0(z) &= -\frac{1}{3}\Lambda_0 z ^{2} \bigg[ 1- \bigg(\frac{z_0}{z}\bigg)^3\bigg],
\end{align}
where the aforementioned horizon is then given by
%
\begin{align}
z_0^3 &= -\frac{12 M_0}{\Lambda_0 L}. 
\end{align}
Due to the cubic nature of the line element there are
three possible horizons, however, only one of them is real and it is defined as the classical 
event horizon. 
The two extra imaginary roots of (\ref{f0cl}) have no physical meaning.
Notice that $M_0$ is the classical black hole mass. What is more, given the noncompactibility of the coordinates $x$ and $y$, we only consider the mass per unit area in the $x-y$ plane \cite{Cai:1996eg}.

At this point we move to thermodynamics of the black plane solutions. 
The starting point is the Euclidean action method \cite{Cai:1994np,Gibbons:1976ue}. 
First, note that the metric can be written in terms of the Euclidean time $\tau$ after the change $t \rightarrow -i \tau$
\begin{align}\label{beta}
\mathrm{d}s^2 &= f_0(z) \mathrm{d}\tau^2 + f_0(z)^{-1}\mathrm{d}z^2 + (L z)^2(\mathrm{d}x^2 + \mathrm{d}y^2).
\end{align}
Thus, in order to obtain the Hawking temperature, the requirement of the absence of the conical singularity in the Euclidean space-time \eqref{beta} causes the Euclidean time $\tau$ to have a period $\beta_0$, which verifies that the temperature is given by
\bea \nonumber
T_0(z_0) &=& \frac{1}{4 \pi} 
\Bigg| 
\lim_{z \rightarrow z_0} \frac{\partial_z g_{tt}}{\sqrt{-g_{tt}g_{rr}}} \Bigg| \\
&=&
\frac{1}{2\pi} \Bigg|
\Bigg(\frac{3 \Lambda_0^2}{L}\Bigg)^{1/3}
\Bigg(\frac{1}{2}M_0\Bigg)^{1/3}
\Bigg|
\ \propto \ M_0^{1/3}.
\eea
%
%
%
%
Following the same line, the Bekenstein--Hawking entropy is given by the usual relation
%
\begin{align}
S_0(z_0) &=  \frac{1}{4} \sigma_0 = 
\frac{3}{2} \pi \Bigg(\frac{L}{3 \Lambda_0^2}\Bigg)^{1/3}  \Bigl( 4 M_0\Bigl)^{2/3} 
\ \propto \ M_0^{2/3},
\end{align}
where the area of the horizon $\sigma_0$, per unit length, is in this case 
given by
\begin{align}
\sigma_{0} &= 2 \pi L z_0^2.
\end{align}
Finally, The heat capacity is 
\begin{align}
C_0(z_0) &= T \ \frac{\partial S}{\partial T} \ \Bigg{|}_{z_0} =-S_0.
\end{align}
It is remarkable that the temperature goes as $M_0^{1/3}$, which strongly differs from the Schwarzschild black hole 
\cite{Frolov:1998wf}. In this sense, both the negative cosmological constant and the planar topology of the horizon
introduce a strong deviation from the Schwarzschild black hole case.
In addition, it should be noted that, when $3L^2 = -\Lambda_0$, the aforementioned results are precisely those 
given in Ref. \cite{Fatima:2011dr}.

\section{Scale--dependent gravity}\label{scale_setting}

As was previously commented in the introduction, one possible way of obtaining a self--consistent theory beyond 
General Relativity is, roughly speaking, by promoting the classical coupling constants to scale--dependent quantities. 
In this sense, effective quantum corrections to well--known black hole solutions
are typically incorporated in two different 
ways: 
i) starting from the effective action, we vary $\Gamma[g_{\mu \nu},k]$ to obtain the effective Einstein equations, 
and ii) starting from the solution, we replace the classical couplings with scale--dependent couplings. 
In particular, we focus on the first situation. The purpose of this section is to summarize the equations of motion for the 
scale--dependent Anti-de Sitter theory. 
Along this paper, we will follow the idea and notation adopted in Ref. \cite{Rincon:2017ypd,Rincon:2017ayr,Contreras:2017eza,
Contreras:2018dhs,Koch:2014joa,Hernandez-Arboleda:2018qdo,Contreras:2018swc,
Rincon:2018lyd,Rincon:2018sgd}. 
After recognizing both the scale--dependent couplings of the theory, which are the
Newton's coupling $G_k$ (which can be related with the
gravitational coupling by $\kappa_k \equiv 8 \pi G_k$), and the cosmological coupling $\Lambda_k$ and the
two independent fields, i.e., the metric field  $g_{\mu \nu}(x)$ and the energy scale $k$,
%
%
the scale--dependent Einstein--Hilbert effective action reads
\begin{eqnarray}\label{action}
\Gamma[g_{\mu\nu},k]=\int \mathrm{d}^{4}x\sqrt{-g}
\bigg[\frac{1}{2 \kappa_{k}} \bigg(R-2\Lambda_{k}\bigg)\bigg],
\end{eqnarray}
where $k$ is a scale-dependent field related to a renormalization scale, and $G_{k}$ and $\Lambda_{k}$ stand for the scale--dependent gravitational and cosmological couplings, respectively.
First, taking variations with respect to the metric field $g_{\mu\nu}$ leads to modified Einstein's equations
\begin{eqnarray}\label{einstein}
G_{\mu\nu}+g_{\mu\nu}\Lambda_{k}=-\Delta t_{\mu\nu},
\end{eqnarray}
where we use the so--called non--matter energy--momentum tensor, $\Delta t_{\mu\nu}$, defined according to  \cite{Reuter:2003ca,Koch:2010nn}
\begin{eqnarray}\label{nme}
\Delta t_{\mu\nu}=G_{k}\left(g_{\mu\nu}\square -\nabla_{\mu}\nabla_{\nu}\right)G_{k}^{-1}.
\end{eqnarray}

We note that the strength of the gravitational and cosmological couplings, $G_k$ and $\Lambda_{k}$, determine the deviation 
of the theory with respect to the corresponding classical case, as expected.
%
Second, taking the variation of the effective action with respect to the field $k(x)$, one 
imposes~\cite{Koch:2014joa}
\begin{align}\label{scale}
\frac{\mathrm{d}}{\mathrm{d} k} \Gamma[g_{\mu \nu}, k] =0.
\end{align}
This condition can be seen as an a posteriori condition towards background independence
\cite{Stevenson:1981vj,Reuter:2003ca,Becker:2014qya,Dietz:2015owa,Labus:2016lkh,Morris:2016spn,Ohta:2017dsq}.
We must emphasize that the aforementioned equation gives a restriction between $G_k$ and $\Lambda_k$ which 
reveals that the cosmological parameter is indeed required to produce self--consistent scale--dependent solutions, at least when
the matter sector is absent.

However, in order to solve these equations, we need the knowledge of the precise beta functions of the problem. 
Given that, in general, an unique solution for the beta functions is still an open question, we can avoid to assume 
any particular form for those. This means that
we do not have enough information in order to find both $g_{\mu\nu}(x)$ and $k(x)$. One possibility to circumvent this issue is by considering that the couplings $\{G_k$, $\Lambda_k\}$ inherit the dependence on space--time coordinates from the space-time dependence of $k(x)$, thus the couplings are written as $\{G(x)$, $\Lambda(x)\}$
\cite{Koch:2014joa,Rincon:2017goj,Rincon:2017ypd}, in combination
with a simplifying ansatz for the line element. 
Although this procedure allows to solve the problem, a high degree of symmetry is usually necessary in order to obtain exact solutions.

In the next section we shall apply this method in order to obtain planar black hole solutions. 
\\
\section{Scale-dependent planar AdS black hole}\label{BlackHoleSolution}

In order to obtain the complete solution with planar symmetry, we need to find the set $\{G(z), \Lambda(z)\}$. 
The running of the gravitational coupling introduces the tensor $\Delta t_{\mu \nu}$ and the effective Einstein field 
equations are

\begin{equation}
\mathcal{S}_{\mu\nu} \equiv G_{\mu\nu}+g_{\mu\nu}\Lambda(z)+\Delta t_{\mu\nu}=0.    
\end{equation}
The so--called non--matter energy momentum tensor, which encodes the running of the Newton coupling, 
is demanded to be zero in the classical limit. Therefore, a well--defined classical limit for the gravitational coupling 
should be imposed. This is achieved through the integration constants which play a crucial role here.
%
Now we will move to the line element used to properly describe the geometry of this problem. 
Specifically, we will consider the line element parametrized as 
\begin{align}\label{Metric}
\mathrm{d}s^2 &= -f(z) \mathrm{d}t^2 + f(z)^{-1}\mathrm{d}z^2 + \left( L z \right)^2(\mathrm{d}x^2 + \mathrm{d}y^2),
\end{align}
where both $\Lambda$ and $G$ depend only on the $z$--coordinate due to the
planar symmetry.
First, the scale--dependent gravitational coupling solving one of the gravitational field
equations has to obey
\begin{align}
G(z)\frac{\mathrm{d}^{2}G(z)}{\mathrm{d}z^{2}} - 2\left(\frac{\mathrm{d} G(z)}{\mathrm{d}z}\right)^2=0,
\end{align}
which allows us to obtain the now well-know scale--dependent solution
\begin{align} \label{Gsol}
G(z) &= \frac{G_0}{1 + \epsilon z},
\end{align}
where $\epsilon$ controls the intensity of the running of the gravitational coupling. 
%
%
%
%
%
%
%
%
The rest of the field equations allow us to find the solution for the lapse function, which we write as 
\begin{align}
f(z) &= f_0(z) + \frac{6 M_0}{L}\epsilon Y(z)
\end{align}
where the auxiliary function $Y(z)$ is defined to be
\begin{align}
Y(z) & \equiv 1 - 2 \epsilon z + 2 (\epsilon z)^2 \ln \left(1 + \frac{1}{\epsilon z} \right).
\end{align}
Finally, the cosmological scale--dependent coupling is obtained when the corresponding algebraic equation is used, 
which gives
\begin{align}
\begin{split}
\Lambda(z) & = \ \Lambda_0 
+ 
\epsilon
\Bigg( \frac{1}{L z (1 + \epsilon z)^2} \Bigg) 
\lambda(z),
\end{split}
\end{align}
where we have defined another supplementary function, $\lambda(z)$, written as
\begin{align}
\begin{split}
\lambda(z) &= 
\Lambda_0  L z^2 (1 + \epsilon z ) 
+
6 M_0 \epsilon  (1 + 12 z \epsilon  (1 + \epsilon z))
\\
&
- 
36 M_0 z \epsilon ^2 (z \epsilon +1) (2 z \epsilon +1) \ln \left(\frac{1}{z \epsilon }+1\right).
\end{split}
\end{align}
%
%

\begin{figure*}[t!]
\centering
\includegraphics[width=0.49\textwidth]{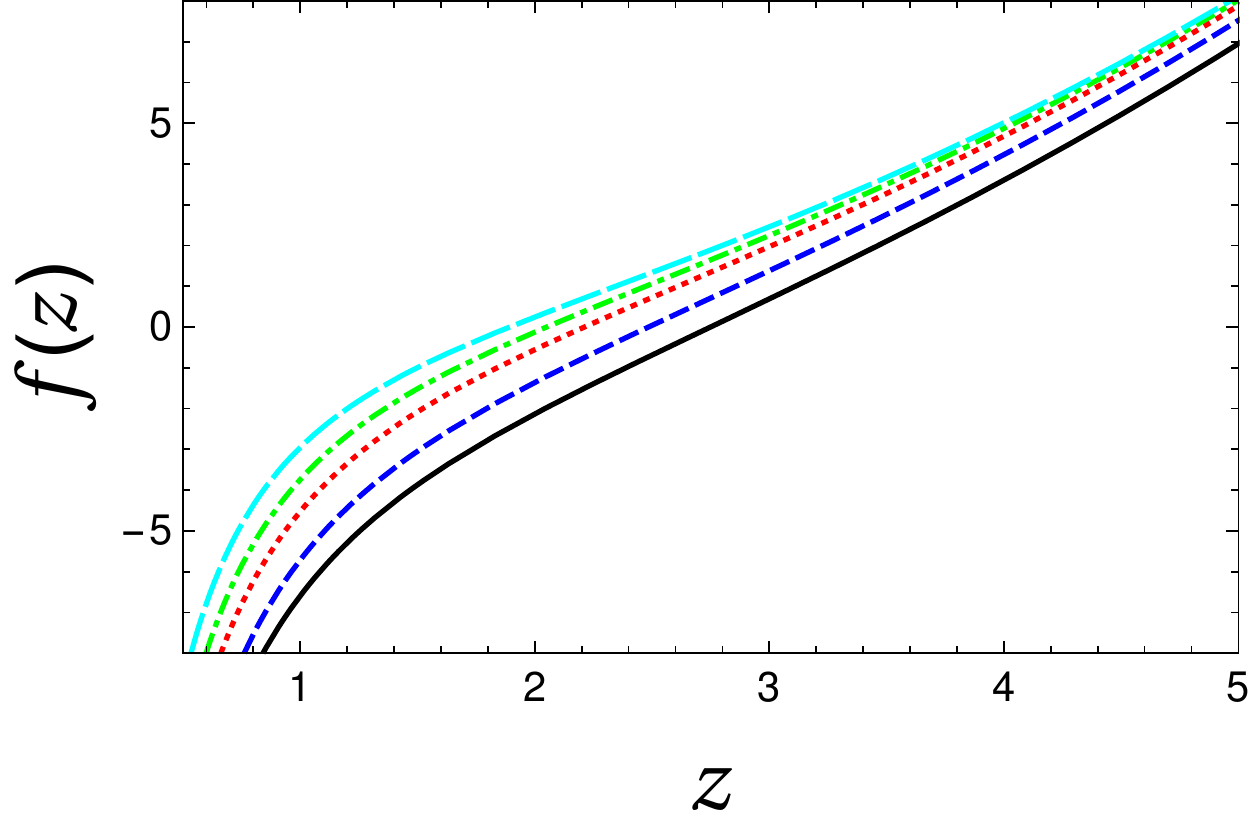} \
\includegraphics[width=0.49\textwidth]{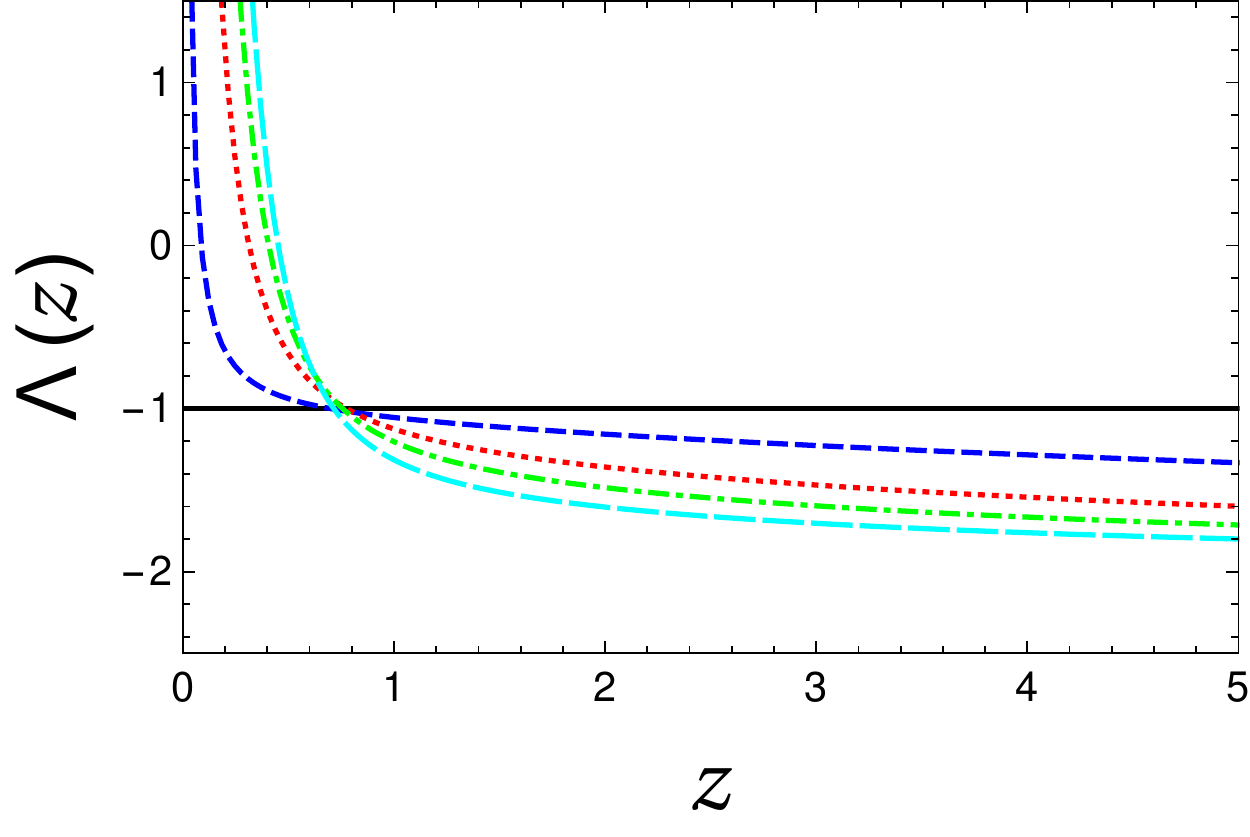}
\caption{
{\bf Left panel:} 
Lapse function $f(z)$ versus $z$ for different values of the running parameter $\epsilon$. 
{\bf Right panel:} Cosmological function $\Lambda(z)$ versus $z$ for different values of the running parameter $\epsilon$. 
The color code correspond: 
i)   $\epsilon =0.0$ (solid black line)
ii)  $\epsilon =0.1$ (short dashed blue line)
iii) $\epsilon =0.3$ (dotted red line)
iv)  $\epsilon =0.5$ (dotted-dashed green line)
v)   $\epsilon =0.8$ (long dashed cyan line)
The classical mass $M_0=1$, the parameters $L = 1/\sqrt{3}$ and $\Lambda_0=-1$ were used in the aforementioned figures.
}

\label{fig:1}
\end{figure*}

Note that the integration constants have been chosen such that we recover the classical solution after turning off 
the running parameter in the functions involved, as can be revealed in Fig. (\ref{fig:1}). Specifically, 
taking $\epsilon\to0$ in the scale--dependent solution we recover
\begin{align}
\lim_{\epsilon \rightarrow 0} G(z) &= G_0 = 1,
\\
\lim_{\epsilon \rightarrow 0} f(z) &= f_0(z) = \bigl(L z \bigl)^{2} \bigg[ 1- \bigg(\frac{z_0}{z}\bigg)^3\bigg],
\\
\lim_{\epsilon \rightarrow 0} \Lambda(z) &= \Lambda_0.
\end{align}
Even more, 
as in general the scale--dependent effects are assumed to be weak, the running parameter is assumed to be small with
respect to the other scales entering the problem such as $M_0$ and $G_{0}$ \cite{Koch:2014joa}, we can write 
\begin{align}
G(z) & \approx G_0(1 - \epsilon z )+ \mathcal{O}(\epsilon^2),
\\
f(z) & \approx f_0(z) + \frac{6 M_0}{L} \epsilon 
+ 
\mathcal{O}(\epsilon^2), 
\label{f_small_epsilon}
\\
\Lambda(z) & \approx \Lambda_0 (1 + \epsilon z) + \mathcal{O}(\epsilon^2).
\end{align}
Interestingly,
as pointed out in \cite{Koch:2014joa} regarding other scale--dependent geometries, the solution here employed 
reveals novel long-range effects due to the scale--dependence because, for $\epsilon \to 0$, the
coordinate $z$ would have to be very large in order to note a deviation from the classical solution. 
Within this limit we have
%
%
\begin{align}
f(z) &=  -\frac{1}{3}\Lambda_0 z^2 - \frac{3 M_0}{L \epsilon z^2} + \mathcal{O}{(z^{-3})} ,
\end{align}
%
which indicates that the AdS radius is not modified, in contrast with \cite{Koch:2014joa}, but an
effective electric charge appears when $\epsilon<0$, as can be shown due to the planar charged black hole $z^{-2}$
dependence for the lapse function.  
Finally, we note that the singular behaviour at $z\to 0$ persists, as a straightforward computation of the curvature 
invariants reveals.

\section{Invariants and Thermodynamics}\label{IT}

A useful way of exploring possible problems in a black hole solution is to investigate the corresponding invariants of 
the geometry. In principle, they can reveal if any problem arises in certain sector of the solution. For instance, the Ricci scalar for 
the classical black hole solution is given in terms of the lapse function as:
\begin{align}
R_0 &= -f_0''(z) -\frac{4 f_0'(z)}{z}-\frac{2 f_0(z)}{z^2},
\end{align}
and it turns out that $R_0= 4 \Lambda_0$ is a constant for any value of $z$. 
In contrast, in the scale--dependent scenario, the Ricci scalar becomes 
extremely complicated and indeed we observe that the $z=0$ singularity, which was already present in the Kretschmann scalar for the
classical solution, now appears also in the Ricci scalar. 
This characteristic is intrinsically related to our formalism and, 
as far as we known, cannot be avoided. 

\begin{figure*}[t!]
\centering
\includegraphics[width=0.49\textwidth]{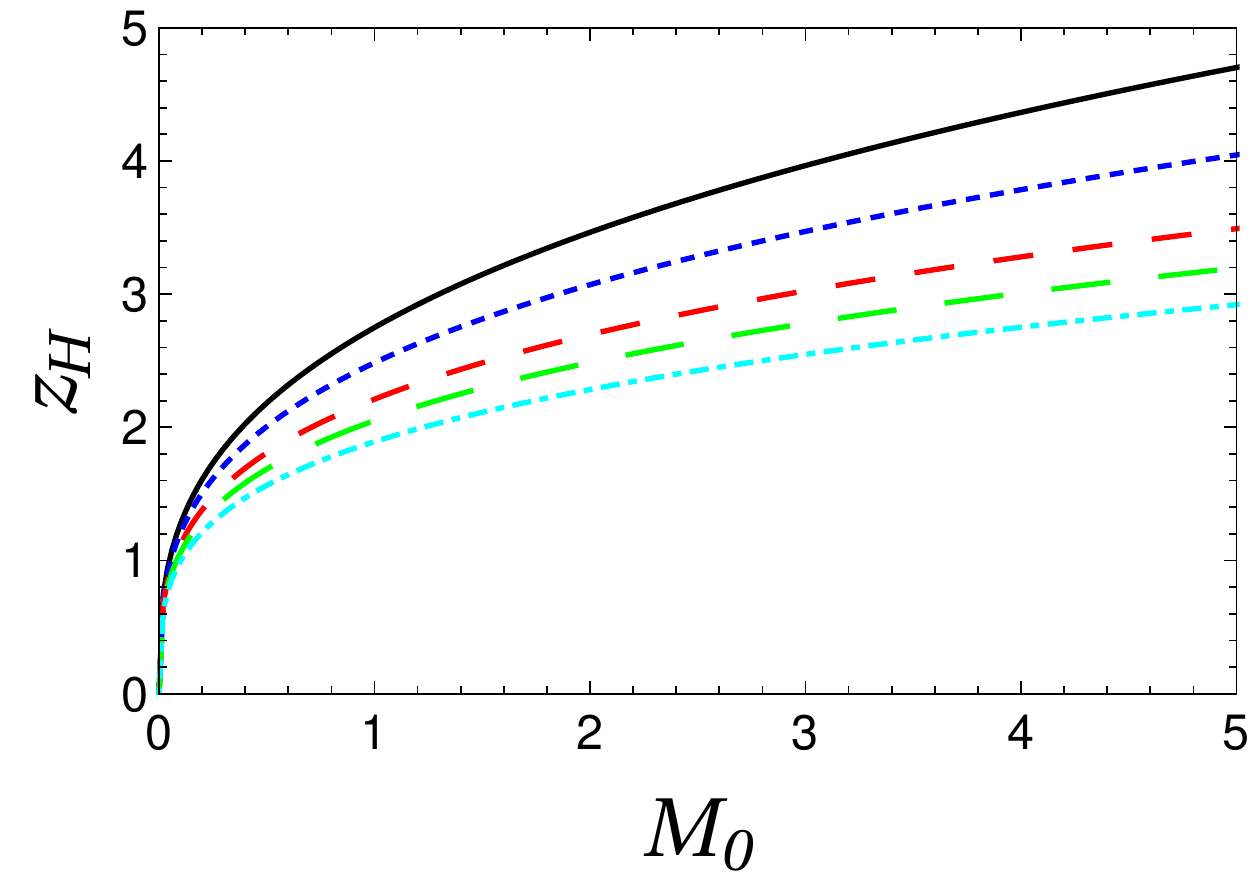} \
\includegraphics[width=0.49\textwidth]{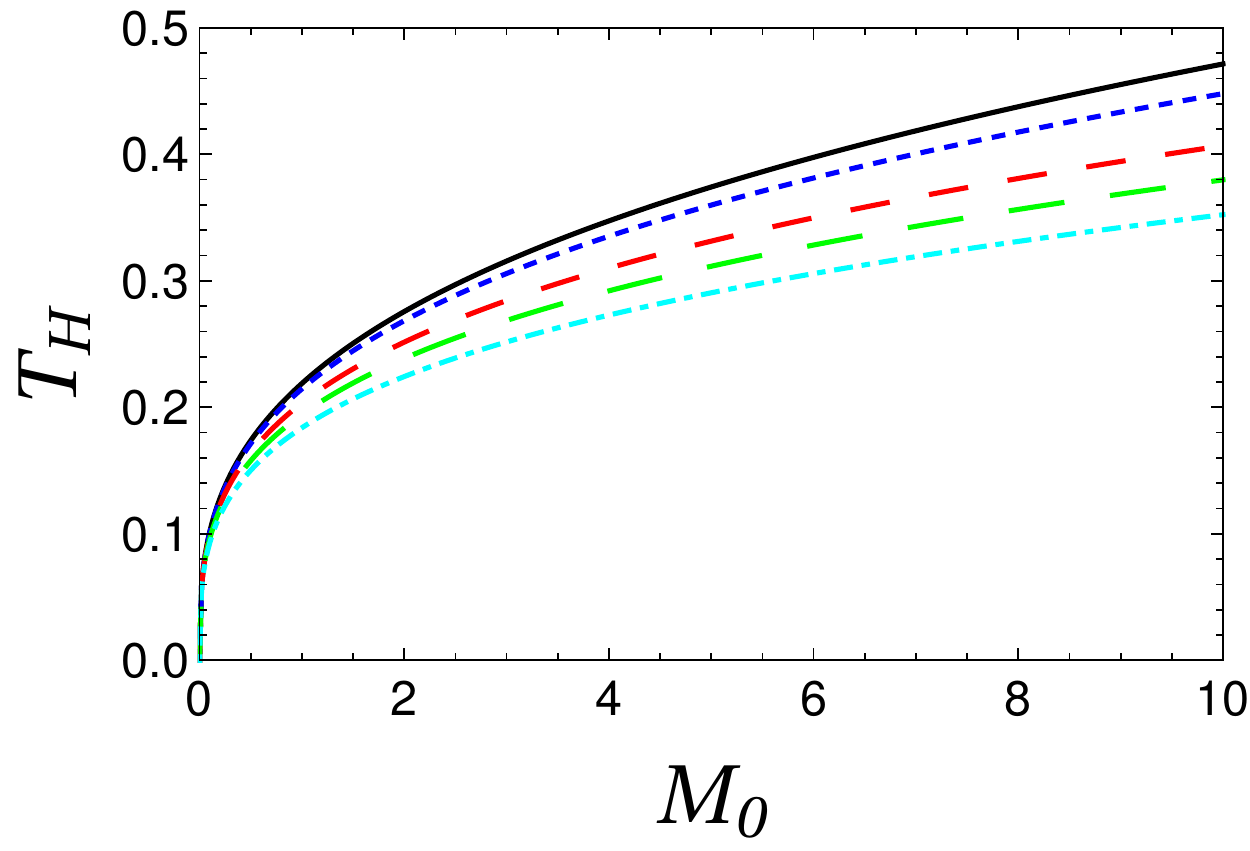}
\caption{
{\bf Left panel:} 
Horizon $z_H$ versus $M_0$ for different values of the running parameter $\epsilon$. 
{\bf Right panel:} Hawking temperature $T_H$ versus $M_0$ for different values of the running parameter $\epsilon$. 
The color code corresponds to: 
i)   $\epsilon =0.0$ (solid black line)
ii)  $\epsilon =0.1$ (short dashed blue line)
iii) $\epsilon =0.3$ (dashed red line)
iv)  $\epsilon =0.5$ (long dashed green line)
v)   $\epsilon =0.8$ (dotted dashed cyan line)
The classical mass has been taken as $M_0=1$, while the parameters $L = 1/\sqrt{3}$ and $\Lambda_0=-1$ were used in the aforementioned 
	figures.
}
\label{fig:2}
\end{figure*}

Before analyzing the thermodynamics, we must focus our attention on the horizon radius. 
In this case, the event horizon is obtained using the condition $f(z_H) = 0$. In general, the task of obtaining an exact horizon is not always possible. 
This is our case because there is a logarithmic contribution to the line element. Still, 
we can obtain a numerical solution for the event horizon and, 
using that, we can analyse the effect of scale--dependent couplings on the AdS planar black hole solution. Moreover, we still can make 
some progress if we take advantage of the small parameter $\epsilon$. As has been mentioned, any deviation with respect to 
the classical solution should be very small, reason why we can assume that $\epsilon$ small provide us an acceptable solution.
Thus, using the approximation given by Eq. \eqref{f_small_epsilon} we obtain
\begin{align}
z_H &= z_0 \left(1 - \frac{1}{2}\epsilon z_0 \right) + \mathcal{O}(\epsilon^2),
\end{align}
where we again observe that the horizon is smaller that the one corresponding to the classical AdS planar solution. Note that this can
also be shown in Fig. \ref{fig:2} (left). 

Although it is important to note the appearance of a shifted horizon with respect to its classical counterpart, 
we do not expect substantial deviations from classical black hole thermodynamics since, as commented previously, 
only long--range effects might show scale--dependent modifications in an appreciable way. In this sense, black hole thermodynamics 
remains robust 
\cite{Contreras:2013hua,Koch:2013rwa,Rodrigues:2015hba,Koch:2015nva,Koch:2016uso,Rincon:2017goj,Rincon:2017ypd,Rincon:2018sgd,Contreras:2018dhs,Rincon:2018dsq,Contreras:2018gct,Contreras:2018gpl,Rincon:2018lyd,Hernandez-Arboleda:2018qdo,Contreras:2018swc}.

%

\begin{figure*}[t!]
\centering
\includegraphics[width=0.49\textwidth]{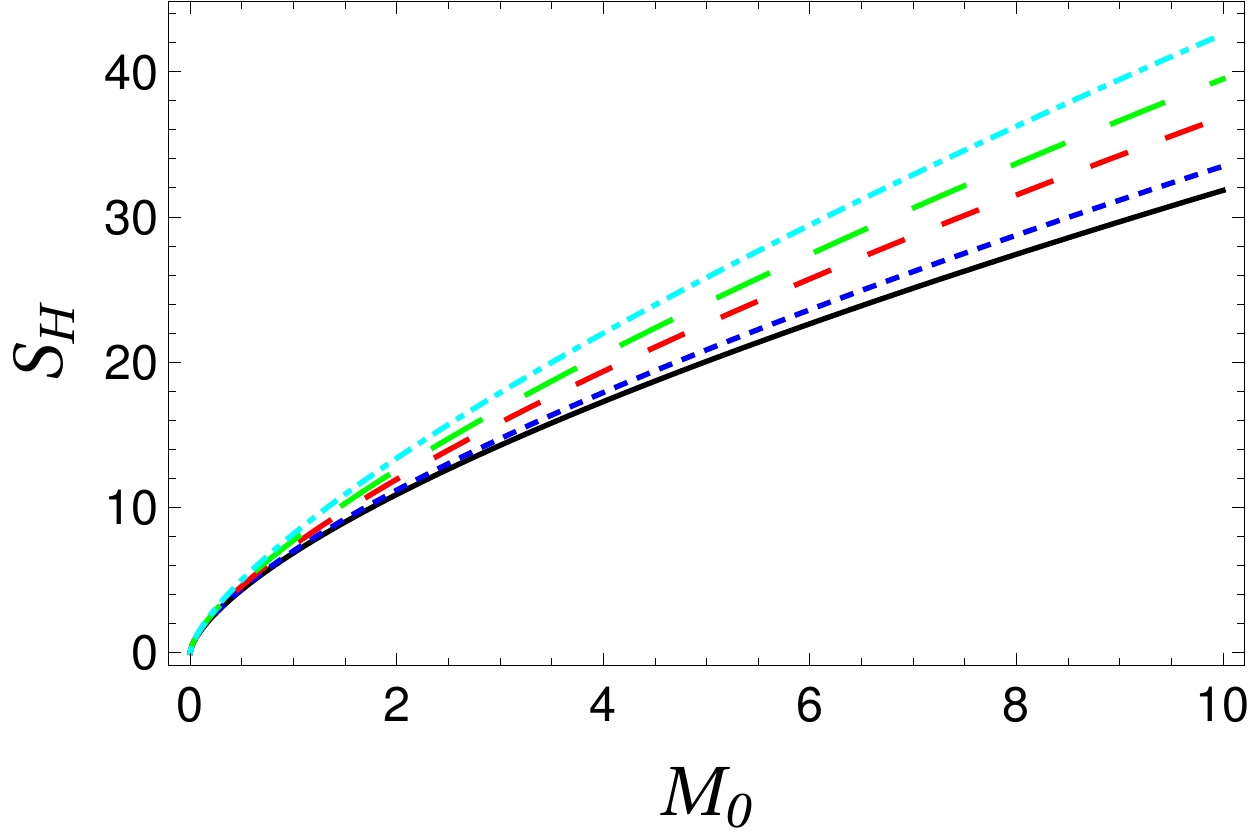} \
\includegraphics[width=0.49\textwidth]{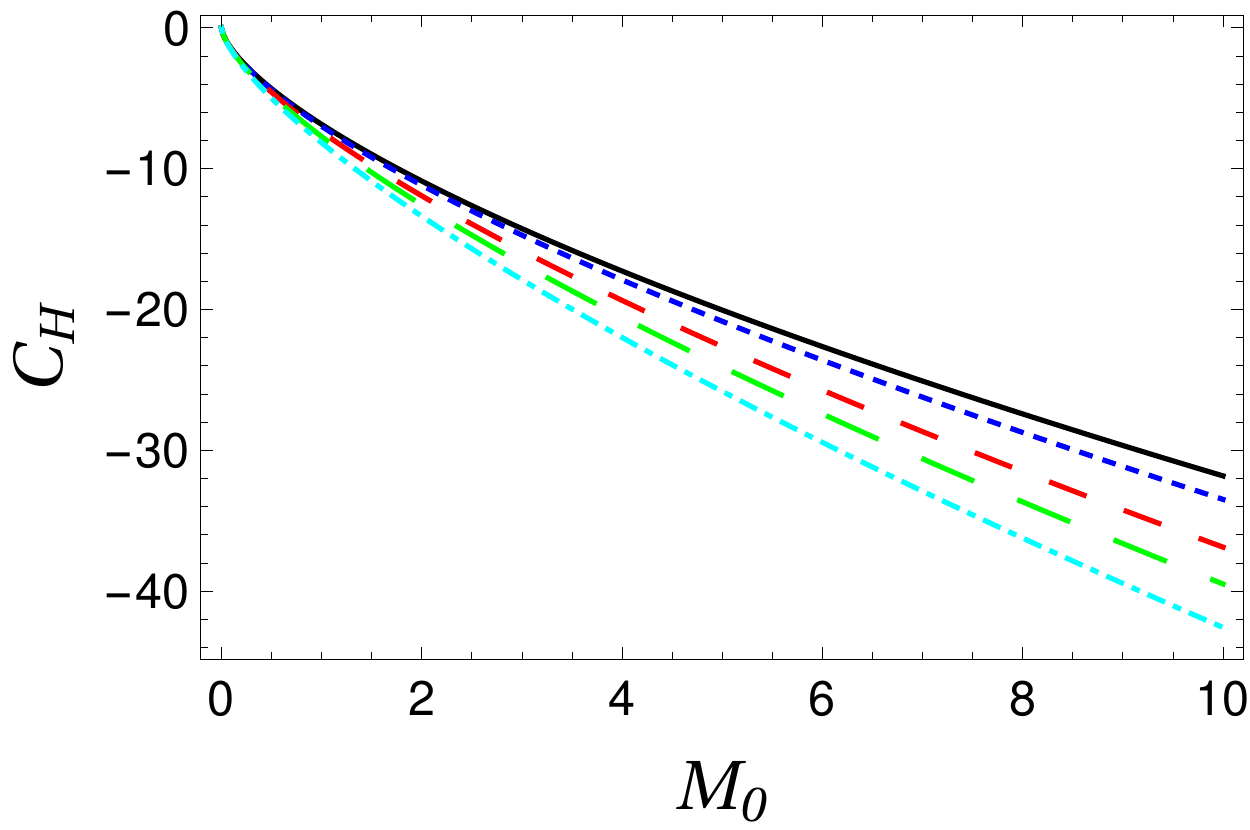}
\caption{
{\bf Left panel:} 
Bekenstein Hawking entropy $S_H$ versus $M_0$ for different values of the running parameter $\epsilon$. 
{\bf Right panel:} Heat capacity $C_H$ versus $M_0$ for different values of the running parameter $\epsilon$. 
The color code correspond: 
i)   $\epsilon =0.0$ (solid black line)
ii)  $\epsilon =0.1$ (short dashed blue line)
iii) $\epsilon =0.3$ (dashed red line)
iv)  $\epsilon =0.5$ (long dashed green line)
v)   $\epsilon =0.8$ (dotted dashed cyan line)
The classical mass $M_0=1$, the parameters $L = 1/\sqrt{3}$ and $\Lambda_0=-1$ were used in the aforementioned figures.
}
\label{fig:3}
\end{figure*}

Regarding black hole thermodynamics, some comments are in order. First, the Hawking temperature is given by
\begin{align}
T_H(z_H) &= 
\frac{1}{4 \pi} 
\Bigg| 
\frac{12 M_0}{L z_H^2 (1 + \epsilon z_H )}
\Bigg|, 
\end{align}
showing that it has a correction via the scale--dependent gravitational coupling. When we demand $\epsilon \rightarrow 0$, the standard 
solution is, of course, recovered.
In order to get some insight about how the scale--dependent scenario affects the temperature with respect to the classical solution, 
we expand for small values of $\epsilon$ to obtain
\begin{align}
T_H(z_H) &= T_0(z_0) \left(1 - \frac{3}{4} (\epsilon z_0)^2 \right) + \mathcal{O}(\epsilon^3).
\end{align}
The previous expression reveals that the temperature decreases with respect to the classical case, $T_{0}(z_{0})$, which is in agreement with the
behaviour shown in Fig. \ref{fig:2} (right).
Second, the Bekenstein-Hawking entropy have the well-known relation inherited from Brans-Dicke theory 
\cite{Kang:1996rj} which, in 3+1 dimensions, reads
\begin{align}
S(z_H) = S_0(z_H) (1 + \epsilon z_H)
\end{align}
and this quantity, as opposed to the temperature, increases when $\epsilon > 0$ and decreases when $\epsilon < 0$. 
It is thus remarkable that, although the expression for entropy admits both positive and negative values for the parameter $\epsilon$, 
we must be careful since $S$ must be positive.
Therefore, this could be considered a point against considering negative values for $\epsilon$.
Finally, the heat capacity is easily computed with help of the relation
\begin{align}
C_H(z_H) &= T \ \frac{\mathrm{d}S}{\mathrm{d}T} \bigg|_{z_H} = -S_H(z_H),
\end{align}
where we notice that always $C_H < 0$, which means that the black hole is indeed unstable. We shown the entropy 
in Fig. \ref{fig:3} (left) and the heat capacity in Fig. \ref{fig:3} (right) for different values of the running parameter $\epsilon$. 
In these figures we can see that the scale--dependent effect is only appreciated when $M_0$ is large.

\section{Concluding remarks}
In this article we have studied the scale dependence
of four dimensional Anti de--Sitter Planar black holes. 
After presenting the model and the classical black hole solution, we have allowed for a scale dependence of the cosmological as well as the gravitational coupling, and we have solved the corresponding
generalized field equations in four-dimensional spacetimes with
planar symmetry. We have analysed in detail some black hole properties such as horizon structure, Hawking
temperature, Bekenstein-Hawking entropy as well
as the heat capacity. In the previous thermodynamics 
quantities we observe that the running correction appears when $M_0$ is large, opposite to what is usually found in 
solutions based on the asymptotic safety program. 

If one compares our result for the running gravitational coupling with the
corresponding results provided by the AS program \cite{Wetterich:1992yh,Morris:1993qb,Reuter:1996cp,Reuter:2001ag,Litim:2002xm,Litim:2003vp,Niedermaier:2006wt,Niedermaier:2006ns,Gies:2006wv,Machado:2007ea,Percacci:2007sz,Codello:2008vh,Benedetti:2009rx,Manrique:2009uh,Manrique:2010am,Manrique:2010mq,Eichhorn:2010tb,Litim:2011cp,Falls:2013bv,Dona:2013qba,Falls:2014tra,Eichhorn:2018yfc,Eichhorn:2017egq} one finds that a
matching is straight forward for  the scale setting choice  $k(z)\sim z$.
This choice seems peculiar, since one usually expects $k \sim 1/z$ for dimensional reasons.
Similar results have been found in \cite{Contreras:2013hua,Koch:2013rwa,Rodrigues:2015hba,Koch:2015nva,Koch:2016uso,Rincon:2017goj,Rincon:2017ypd,Rincon:2018sgd,Contreras:2018dhs,Rincon:2018dsq,Contreras:2018gct,Contreras:2018gpl,Rincon:2018lyd,Contreras:2018swc}
but the deeper reason behind this result is still unknown .
An important hint for solving this riddle could come from considering the dimensionless product
$G(k)\cdot \Lambda(k)$ instead of the individual dimensionful quantities as discussed in~\cite{Canales:2018tbn}.

Another interesting feature of our solution is that
the event horizon is attenuated in the scale--dependent scenario, which means that the black hole is smaller that the classical solution. Regarding the temperature, we notice that it is lower than in the classical case, whereas the entropy is larger than
that of the non-running case. Finally, we have noted that the heat capacity is negative, which implies that the black hole is 
unstable. All these features give a better comprehension of the effect of scale--dependent couplings in well known black hole solutions.

\section*{Acknowledgments}
The author A. R. was supported by the  CONICYT-PCHA/Doctorado  Nacional/2015-21151658. The author B. K. was supported by  
Fondecyt 1161150 and Fondecyt 1181694. 
The author P. B. was supported by the Faculty of Science and Vicerrector\'{\i}a de Investigaciones of Universidad de los Andes, Bogot\'a, Colombia.

\end{document}